\documentclass[aps,prapplied,reprint,superscriptaddress]{revtex4-2}

\usepackage{graphicx}
\usepackage[uncertainty-mode = separate, separate-uncertainty-units = single,  bracket-unit-denominator = false]{siunitx}
\usepackage{textgreek}
\usepackage[hidelinks]{hyperref}

\usepackage[utf8]{inputenc}
\usepackage[T1]{fontenc}
\usepackage{tabularx}

\begin{document}

\title{Efficient fiber-pigtailed source of indistinguishable single photons}

\author{Nico Margaria}
\affiliation{Quandela, 91300 Massy, France}
\author{Florian Pastier}
\affiliation{Quandela, 91300 Massy, France}
\author{Thinhinane Bennour}
\affiliation{Quandela, 91300 Massy, France}
\author{Marie Billard}
\affiliation{Quandela, 91300 Massy, France}
\author{Edouard Ivanov}
\affiliation{Quandela, 91300 Massy, France}
\author{William Hease}
\affiliation{Quandela, 91300 Massy, France}
\author{Petr Stepanov}
\affiliation{Quandela, 91300 Massy, France}
\author{Albert F. Adiyatullin}
\affiliation{Quandela, 91300 Massy, France}
\author{Raksha Singla}
\affiliation{Quandela, 91300 Massy, France}
\author{Mathias Pont}
\affiliation{Quandela, 91300 Massy, France}
\author{Maxime Descampeaux}
\affiliation{Quandela, 91300 Massy, France}
\author{Alice Bernard}
\affiliation{Quandela, 91300 Massy, France}
\author{Anton Pishchagin}
\affiliation{Quandela, 91300 Massy, France}
\author{Martina Morassi}
\affiliation{Université Paris-Saclay, CNRS, Centre de Nanosciences et de Nanotechnologies, 91120 Palaiseau, France}
\author{Aristide Lema\^itre}
\affiliation{Université Paris-Saclay, CNRS, Centre de Nanosciences et de Nanotechnologies, 91120 Palaiseau, France}
\author{Thomas Volz}
\affiliation{Quandela GmbH, 81671 Munich, Germany}
\author{Valérian Giesz}
\affiliation{Quandela, 91300 Massy, France}
\author{Niccolo Somaschi}
\email{niccolo.somaschi@quandela.com}
\affiliation{Quandela, 91300 Massy, France}
\author{Nicolas Maring}
\affiliation{Quandela, 91300 Massy, France}
\author{Sébastien Boissier}
\affiliation{Quandela, 91300 Massy, France}
\author{Thi Huong Au}
\affiliation{Quandela, 91300 Massy, France}
\author{Pascale Senellart}
\affiliation{Université Paris-Saclay, CNRS, Centre de Nanosciences et de Nanotechnologies, 91120 Palaiseau, France}

\date{\today}

\begin{abstract} 
Semiconductor quantum dots in microcavities are an excellent platform for the efficient generation of indistinguishable single photons.
However, their use in a wide range of quantum technologies requires their controlled fabrication and integration in compact closed-cycle cryocoolers, with a key challenge being the efficient and stable extraction of the single photons into a single-mode fiber.
Here we report on a novel method for fiber-pigtailing of deterministically fabricated single-photon sources.
Our technique allows for nanometer-scale alignment accuracy between the source and a fiber, alignment that persists all the way from room temperature to \qty{2.4}{\kelvin}.
We demonstrate high performance of the device under near-resonant optical excitation with g\textsuperscript{(2)}(0)~=~\qty{1.3}{\percent}, a photon indistinguishability of \qty{97.5}{\percent} and a fibered brightness of \qty{20.8}{\percent}.
We show that the indistinguishability and single-photon rate are stable for over ten hours of continuous operation in a single cooldown.
We further confirm that the device performance is not degraded by nine successive cooldown-warmup cycles.
\end{abstract}

\maketitle

\subsection{Introduction}
Recent advances in quantum photonics have led to the generation, processing, and detection of quantum states of light with increasing size and complexity~\cite{zhong_quantum_2020,madsen_quantum_2022}.
The scalability of these protocols is limited by the challenge of generating identical single-photons with high efficiency.
Single-photon sources based on semiconductor quantum dots (QDs) coupled to microcavities have demonstrated unparalleled efficiencies and single-photon purities~\cite{somaschi_near-optimal_2016, wang_towards_2019, uppu_scalable_2020, tomm_bright_2021, ding_high-efficiency_2023}, with no fundamental compromise to optimize both properties simultaneously.
These devices have enabled fast progress in the field of photonic quantum computing~\cite{wang_high-efficiency_2017, wang_deterministic_2023, cao_photonic_2024, fyrillas_certified_2024, maring_versatile_2024} and photonic quantum communication~\cite{vajner_quantum_2022, takemoto_quantum_2015, basso_basset_quantum_2021, schimpf_quantum_2021,gao_quantum_2022, gyger_metropolitan_2022, basso_basset_daylight_2023, morrison_single-emitter_2023, zahidy_quantum_2024}.
Underpinning these recent successes is the ability to precisely control the coupling of single QDs to microcavities as a way to efficiently funnel the emitted photons into single-mode fibers~\cite{gerard_strong_1999}.

A remaining obstacle to the widespread adoption of QD-based single-photon sources is the need to operate them at cryogenic temperatures around \qty{5}{\kelvin}.
Particularly, the requirements of optical excitation and photon collection with nanometer-scale precision and of long-term stability are challenging.
State-of-the-art demonstrations have so far been obtained in laboratory settings with liquid-helium cryostats~\cite{wang_towards_2019,tomm_bright_2021} or ultra low-vibration closed-cycle cryocoolers~\cite{somaschi_near-optimal_2016,uppu_scalable_2020,ding_high-efficiency_2023}, incorporating bulky optical systems between the source and the collection fiber.
A more robust solution for fiber coupling is needed to enable the operation of these bright single-photon sources in standard closed-cycle cryocoolers that are subject to vibration, in the same vein as the development of fiber-coupled superconducting nanowire-based single-photon detectors~\cite{esmaeil_zadeh_single-photon_2017}.

\begin{figure*}[ht]
    \includegraphics[width=17.2cm]{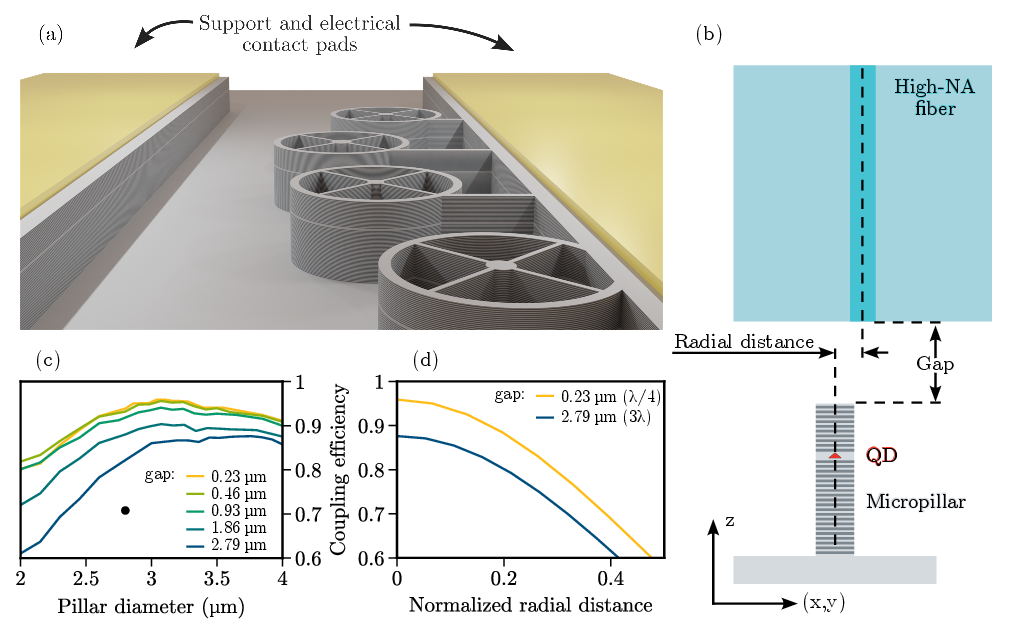}
    \caption{\label{fig1}Direct fiber-coupling of micropillar-based single-photon sources.
    (a) 3D render of the deterministic micropillar devices with the electrical contact pads designed to act as mechanical support in the pigtailed device.
    (b) Illustration of the region of interest around the micropillar and fiber tip, defining important parameters for the alignment of the system.
    (c,d) Numerical simulation of micropillar to high-NA fiber coupling as a function of the gap and of the radial distance normalized to the micropillar radius.
    The black dot in (c) indicate the simulated coupling efficiency for the diameter and gap of our device.}
\end{figure*}

In the past fifteen years, there have been many attempts toward this goal~\cite{bremer_fiber-coupled_2022}, exploiting QDs in photonic wires~\cite{cadeddu_fiber-coupled_2016}, photonic crystal cavities~\cite{shambat_optical_2011, daveau_efficient_2017}, microlenses~\cite{schlehahn_stand-alone_2018,bremer_quantum_2020}, micromesas~\cite{zolnacz_method_2019} or bullseye cavities~\cite{jeon_plug-and-play_2022}.
Most approaches have relied on direct gluing of a fiber on the single-photon device, often with limited mode overlap, thereby significantly reducing the efficiency of the fiber-connected source.
These studies showed high single-photon purity of the extracted photons while few also demonstrated high photon indistinguishability~\cite{snijders_fiber-coupled_2018,rickert_fiber-pigtailed_2024}, a key metric for most quantum applications. 
To achieve high photon indistinguishability, the fiber integration must be combined with electrical control of the single-photon source to reduce charge noise effects.
So far, the most advanced demonstration was reported by Snijders et al.~\cite{snijders_fiber-coupled_2018}, showing a high indistinguishability combined with a fibered brightness of \qty{1.53}{\percent}, defined as the probability of getting a single photon at the output fiber for each laser pulse.
However, these results were obtained with no deterministic placement of the QDs relative to the cavity mode and no active control of the spectral detuning of the QD line with respect to the cavity resonance.
Combining the controlled coupling of the QD to the cavity, the electrical control of the source, and the accurate fiber-pigtailing is highly challenging.
Specifically, the fiber alignment with the device should be controlled at the scale of tens of nanometers and be maintained from room temperature to below \qty{10}{\kelvin}, in an architecture which is robust to many cooling cycles.
In the present work, we report how we have tackled and overcome all these technological challenges, opening the door to a new era of quantum photonic applications based on single QD sources.

Our technology is based on individual semiconductor QDs in micropillar cavities which are directly coupled to single-mode fibers.
We demonstrate the successful integration of our source in a compact two-stage Gifford-McMahon cryocooler operating at a base temperature of \qty{2.4}{\kelvin}.
Our mechanical design is shown to precisely maintain the alignment between the cavity and the fiber during cooldown and is robust enough to handle the vibrations of the cryocooler during operation.
The optical excitation of the source and the photon collection are realized through compact optical modules connected to the pigtailed device.
The source produces a stream of single photons with an emission rate of \qty{16.75(5)}{\mega\hertz} and indistinguishability of \qty{97.5(1)}{\percent}, and is stable over hours of continuous operation, as well as throughout multiple cooling cycles of the cryocooler.

\subsection{Source design and fiber coupling}

Our quantum emitters are single self-assembled InAs QDs grown by molecular beam epitaxy at the center of an unbalanced planar microcavity formed from 20(40)-pairs top(bottom) distributed Bragg reflectors.
The GaAs/Al\textsubscript{0.95}Ga\textsubscript{0.05}As heterostructure is doped with Be and Si to form a vertical p-i-n junction which can be reverse-biased to control the charge state of the QD ground state~\cite{warburton_single_2013} and to tune the emission wavelength by roughly \qty{150}{\pico\metre} through the quantum-confined Stark effect.
The self-assembled QDs are randomly distributed in the plane of the wafer and have an inhomogeneous spectral broadening due to variations in size and composition.
We use the in-situ lithography technique~\cite{dousse_controlled_2008} to precisely locate the QDs on the wafer and fabricate micropillars centered on individual QDs.
We choose the diameter of each micropillar to coarsely match the wavelength of the cavity mode to that of the target QD line (within \qty{200}{\pico\metre}).
For bias voltage tuning, each micropillar on the chip is connected to electrical contacts by four narrow ridges~\cite{nowak_deterministic_2014} (Fig.~\ref{fig1}a).
The number and arrangement of the ridges preserves the Gaussian-like spatial distribution of the fundamental mode.
In addition, the electrical contact pads serve as supporting structures for the fiber pigtailing as described later.
The resulting micropillars have diameters ranging from 2 to \qty{3.5}{\micro\metre} and quality factors of around Q~$\approx$~13000 which leads to a radiative enhancement of the targeted QD emission lines with Purcell factors F\textsubscript{p}~$\approx$~13.
This results in a decay time \texttau~$\approx$~\qty{80}{\pico\second} when the QD emission is tuned into resonance with the cavity using the bias voltage control.
Chips containing dozens of devices are first characterized at \qty{4}{\kelvin} with a free-space optical setup and the best-performing micropillar in terms of single-photon rate, purity, and indistinguishability is identified for each chip~\cite{ollivier_reproducibility_2020}.
They are subsequently brought back to room temperature for the direct fiber-coupling of the chosen micropillars.

For an optimal mode matching between the micropillars and single-mode fibers, we choose a commercially available single-mode fiber (UHNA3) with a numerical aperture (NA) of 0.35 and a core diameter of \qty{1.8}{\micro\metre}.
This fiber is spliced to a standard 780HP fiber for connection with the excitation and collection setup.
The fiber-micropillar system is sketched in Fig.~\ref{fig1}b.
The key parameters are the gap that separates the fiber and the micropillar, and the radial distance between their respective optical axes.
Fig.~\ref{fig1}c presents the results of numerical simulations of the coupling efficiency between the ridge-connected micropillar and the high-NA fiber performed with finite-difference time-domain and mode expansion methods.
The different colors correspond to different gaps between the fiber and the micropillar.
As expected, minimal losses are observed when the gap approaches zero, as there is minimal divergence of the beam in the unguided region.
We find that for a gap size of \qty{0.23}{\micro\metre}, the maximum achievable coupling is around \qty{96}{\percent}.
For a fiber which is perfectly aligned on axis, there is an optimal micropillar diameter which maximizes the mode overlap with the fiber.
We note that the coupling efficiency around the optimal diameter remains above \qty{94}{\percent} over a range of \textpm\qty{0.5}{\micro\metre}, which is larger than the tolerance in the fabricated diameters.
Looking at the curves for the different gap sizes, we estimate the coupling efficiency for the optimal micropillar diameter to decrease to \qty{93.8}{\percent} at \qty{1}{\micro\metre} gap size, and to \qty{89.9}{\percent} at \qty{2}{\micro\metre} gap size.
We find that the optimal diameter increases for larger gaps due to smaller divergence of the light emerging from larger micropillar modes.
Finally, Fig.~\ref{fig1}d presents the coupling efficiency for various radial displacements between the micropillar and fiber core axes.
The results follow the expected convolution between two Gaussian modes evidenced by a slow quadratic variation near zero displacement.
This indicates that the presence of ridges does not significantly alter the Gaussian profile of a cylindrical micropillar cavity.

\begin{figure*}[ht]
    \includegraphics[width=17.2cm]{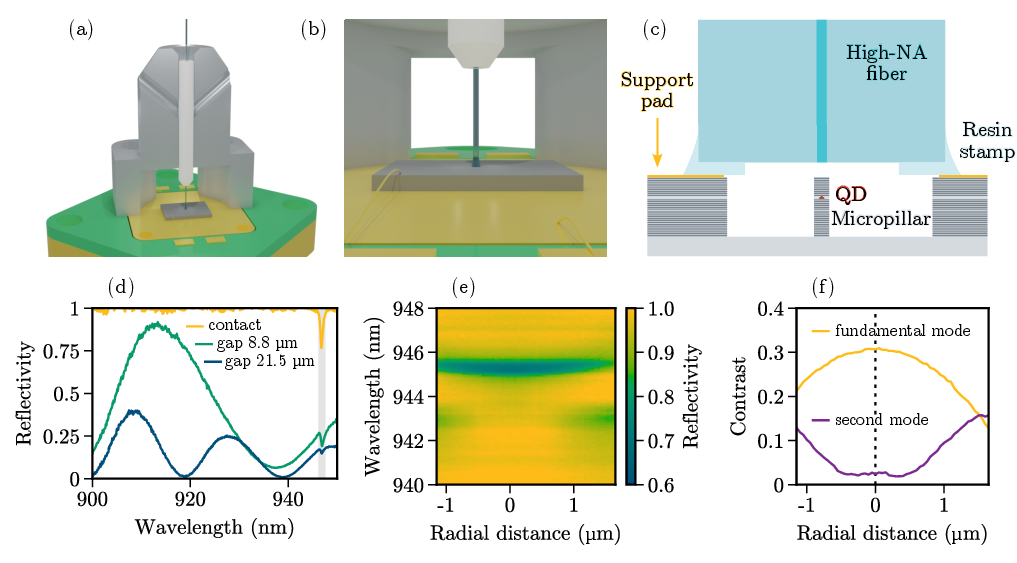}
    \caption{\label{fig2}Fiber alignment procedure.
    (a,b) 3D visualization of the resin stamp and fiber holder (section) for improved stability of the pigtailed system.
    (c) Sketch of the micropillar-fiber section highlighting the contact physical region between the resin stamp and the single-photon emitting device (not in scale).
    (d) Reflectivity spectra of the micropillar at room temperature for different vertical gaps showing broadband modulations from interference effects.
    The fundamental mode spectral region is highlighted.
    (e) 2D color map of the reflectivity spectrum at room temperature across the section of a micropillar.
    (f) Contrast of the two lowest-order modes extracted from (e).}
\end{figure*}

\subsection{Fiber pigtailing}

Our numerical studies imply that direct contact between the micropillar and the fiber is optimal in terms of coupling efficiency.
However, any unwanted source of stress applied by the fiber on the micropillar will lead to wavelength shifts of both the QD emission line and the cavity mode resonance~\cite{moczala-dusanowska_strain-tunable_2019, gerhardt_optomechanical_2020}.
Therefore, we aim for the fiber to be as close as possible to the micropillar without direct contact.
This has to be achieved at cryogenic temperatures after thermal contraction of all the components in the mechanical structure, each of them presenting different thermal contraction coefficients.
To meet these requirements, we employ the structure presented in Fig.~\ref{fig2}a-c.
The fiber, previously glued into a ceramic ferrule, is held in place by a titanium fiber-holder assembly.
To secure the fiber to the chip, we fabricate a customized supporting structure at the end of the fiber which we refer to as the stamp.
The stamp defines a \qty{3}{\micro\metre} gap by design and allows us to push the fiber on top of the electrical contact pads surrounding the micropillar without applying any direct pressure on the micropillar itself.
The mechanical stability obtained in this way is needed to prevent sliding of the fiber tip during cooldown and to eliminate vibrations.
The stamp structure is fabricated by dipping the tip of the fiber in uncured resin, and subsequently shaping the resin by pressing the fiber on a Si-based disk-shaped mold.
To align the fiber core to the mold, a small gold disk is fabricated at the centre of the mold to allow for spatial alignment by using the reflection of a laser back into the fiber.
The resin is then cured by illuminating it with UV light.

To ensure that the fiber is positioned within a few micrometers above the micropillar, a vertical alignment step is first performed to define the relative position of the ferrule in the fiber holder before securing them together.
The ferrule is gradually lowered into the fiber holder and we monitor the reflectivity spectrum resulting from a broadband light source sent through the fiber.
A Fabry-Pérot cavity is formed between the fiber end facet and the device surface, leading to a periodic modulation of the reflectivity spectrum.
The spectra in Fig.~\ref{fig2}d show the reflectivity measured for various vertical distances between the fiber tip and the chip.
The data is normalized with respect to when in direct contact.
The observed broadband modulation of the reflectivity allows us to extract the gap value as $\frac{\text{\textlambda\textsubscript{1}\textlambda\textsubscript{2}}}{\text{2~\textDelta \textlambda}}$, where \textDelta \textlambda ~is the spacing between two successive maxima of the reflectivity spectrum at wavelengths \textlambda\textsubscript{1} and \textlambda\textsubscript{2}.
This technique allows us to precisely measure the vertical distance between the fiber facet and the top of the planar microcavity with a relative accuracy which improves linearly while reducing the distance.
Hence, the distance is continuously monitored as the fiber is lowered in the holder until it lands smoothly on the contact pads.
At this point the ferrule is glued into the fiber-holder.

Once the vertical alignment is done, the fiber holder is separated from the chip and rigidly attached to a precision scanning stage for accurate x, y, and z positioning, thereby maintaining a gap of a few micrometers between the stamp and the surface of the chip.
To position the fiber core precisely above the micropillar center with sub-\unit{\micro\metre} lateral alignment precision, we then exploit specific features in the reflectivity spectrum of the micropillar.
The micropillars support several higher-order transverse modes with resonance wavelengths shorter than the fundamental cavity mode~\cite{gerard_quantum_1996}.
These modes are seen as dips in the reflectivity spectrum whose contrasts depend on their spatial overlap with the Gaussian mode of the fiber.
Coupling to symmetric\,(antisymmetric) modes will be maximized\,(minimized) with perfect centering.
The result of such a measurement is presented in Fig.~\ref{fig2}e, which displays a reflectivity map as a function of wavelength and radial displacement between the micropillar and the fiber core.
We can clearly see the fundamental micropillar mode at \qty{945.8}{\nano\metre} whose contrast is maximized for perfect alignment.
The fundamental mode is also highlighted in Fig.~\ref{fig2}d with a grey vertical bar.
The second transverse mode of the micropillar appears around \qty{943.0}{\nano\metre} with some radial displacement.
To visualize the coupling to the different transverse modes more clearly, we plot the contrast of the fundamental and second order modes dips (defined as the normalized depth of the dip with respect to the background) as a function of radial distance. 
This allows us to accurately align the two structures radially.
From the symmetric behavior of the dip contrast with respect to the lateral displacement, we estimate the alignment precision to be better than \qty{200}{\nano\metre}.
Once the xy-position is aligned, we close the vertical gap between the stamp and the chip, and we secure the fiber holder to the chip holder using screws.
During this final step, the reflectivity spectrum is constantly monitored to check that the transverse alignment is not perturbed.
Furthermore, we underline that our fiber-coupling process is fully reversible.
The fiber holder can be unscrewed, and another micropillar can be targeted without damage nor alteration to the properties of the micropillars on the chip.

\begin{figure}[ht]
    \includegraphics[width=8.6cm]{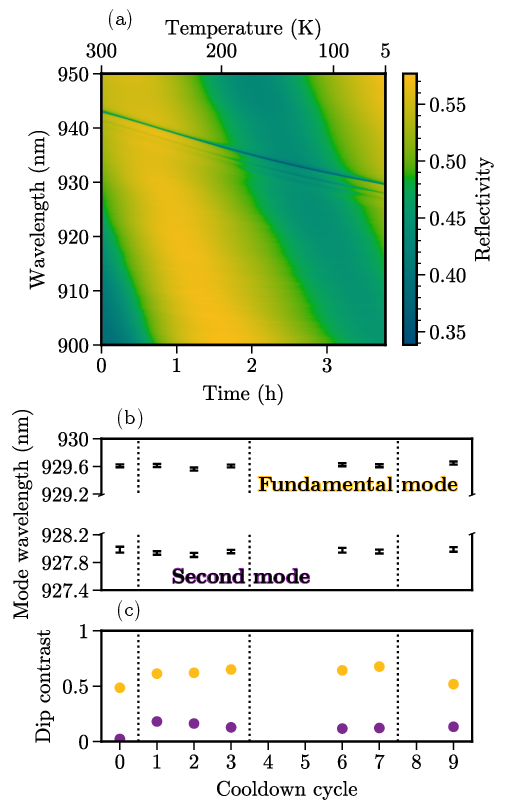}
    \caption{\label{fig3}Stability of direct-fiber coupling.
    (a) Reflectivity monitored during a typical cooldown of the pigtailed single-photon source.
    (b,c) Comparison of reflectivity features (micropillar modes wavelength and contrast) at the cryocooler base temperature in free space and six out of nine successive cooldown cycles in the pigtailed version.
    The dotted lines indicate a change of cryocooler.}
\end{figure}

After completion of the direct coupling process, the fibered source is installed in a compact two-stage Gifford-McMahon cryocooler operating at a base temperature of \qty{2.4}{\kelvin}.
The fiber-holder assembly is in thermal contact with the cold plate of the cryocooler through the chip holder.
To demonstrate the robustness of the alignment, we show the reflectivity spectrum of the pigtailed device as a function of temperature during a typical cooldown in Fig.~\ref{fig3}a, starting from \qty{300}{\kelvin} all the way down to the base temperature of the cryocooler.
The narrow resonance dips correspond to the micropillar modes that are gradually blue-shifting with decreasing temperature due to the change in refractive index of the cavity medium.
More importantly, we do not see the strengthening of higher-order cavity modes at the expense of the fundamental mode, which would indicate a transverse displacement of the fiber with respect to the micropillar (compared to the behavior observed in  Fig.~\ref{fig2}e,f).
We observe a shift in the slow modulation of the background reflectivity due to a contraction of the vertical gap between the fiber and the micropillar (cf. Fig.~\ref{fig2}d).
From the background modulation, we estimate the gap to be \qty{3.5}{\micro\metre}, which is comparable  to the design gap size of \qty{3}{\micro\metre}.
Next, we show in Fig.~\ref{fig3}b,c that the micropillar modes wavelength and contrast are not significantly altered across nine consecutive thermal cycles in three different cryocoolers over several months of operation.
The wavelength fluctuation is less than \qty{30}{\pico\metre} and the second mode contrast is always significantly lower than the fundamental mode one, implying that our technique to hold the fiber achieves remarkable stability with no tangible effect on the microcavity.
We note that the first data point is a reference and was taken during the free-space characterization preceding the pigtailing procedure.
The lower contrast in free-space evidences a different mode matching to the fiber using lenses.

\subsection{Performance of the fiber-pigtailed source}

\begin{figure}[ht]
    \includegraphics[width=8.6cm]{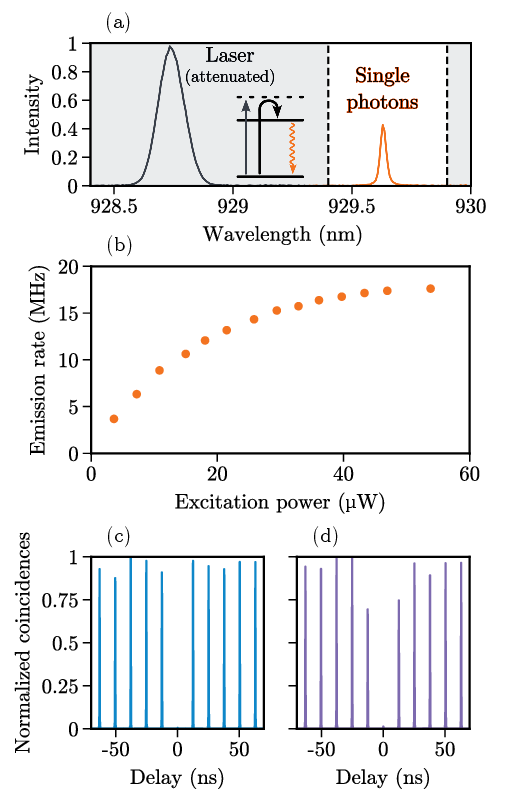}
    \caption{\label{fig4}Characterization of the fiber-coupled single-photon source.
    (a) Spectra of the excitation laser and QD emission; the shaded area is rejected by the filtering in the collection path.
    (b) Emission rate saturation curve as a function of excitation power.
    (c) Second-order auto-correlation function and
    (d) Hong-Ou-Mandel interference measurements.}
\end{figure}

We now present the characterization of the pigtailed source and compare its properties to those measured before fiber integration.
We operate the fibered device with a pulsed near-resonant laser relying on longitudinal acoustic (LA) phonon-assisted excitation~\cite{ardelt_dissipative_2014, quilter_phonon-assisted_2015, reindl_highly_2019, thomas_bright_2021}.
In this excitation scheme, represented in Fig.~\ref{fig4}a, the laser wavelength is blue-detuned from the QD resonance and can be straightforwardly separated from the emitted single photons using spectral filters.
This technique is very robust to fluctuations in the laser wavelength and intensity, and allows us to address a single exciton dipole, leading to highly polarized single photon emission.

We coarsely bias the p-i-n junction to activate the neutral exciton state of the QD and then fine-tune the line in resonance with the cavity mode.
We then optimize the duration of the laser pulses, the laser detuning, and the polarization in order to maximize the performance of the source.
We find the best settings for a pulse duration of \texttau~=~\qty{10}{\pico\second} at a repetition rate of \qty{79.21}{\mega\hertz} and a detuning of \qty{-0.8}{\nano\metre} with respect to the QD emission wavelength.
Finally, the polarization of the laser is finely adjusted to excite only one of two orthogonal dipoles of the neutral exciton.
The emitted photons are thus highly linearly polarized with the degree of polarization reaching \qty{95}{\percent}.

We measure the single-photon rate at the system output and correct the recorded rate for the efficiency of the detector.
Fig.~\ref{fig4}b represents the output single-photon rate as a function of the time-averaged excitation power.
As expected with phonon-assisted excitation, the rate saturates at high excitation power, here at a value of \qty{17.60}{\mega\hertz}.
For the best performance in terms of single-photon purity and indistinguishability, we operate the source slightly below saturation, at a single-photon rate of \qty{16.75}{\mega\hertz}.
This minimizes the fraction of spurious laser light collected while maintaining a high single-photon source efficiency.
From the output single-photon rate and the laser repetition rate, we calculate a fibered brightness of \qty{20.8(8)}{\percent}.
For comparison, the best achieved result so far for pigtailed single-photon sources is a fibered brightness of \qty{5.8}{\percent}~\cite{cadeddu_fiber-coupled_2016}.

We use these results to quantify the micropillar-to-fiber coupling efficiency, providing a direct measure of the fiber-pigtailing process quality.
To access the losses which are solely due to the imperfect coupling from the micropillar fundamental mode into the fiber, we use results from the characterization of the source performed before the fiber-coupling process.
In our free-space characterization setup, we estimate the first lens brightness of our device, defined as the probability of collecting a photon per excitation pulse at the output of the micropillar in a 0.7\,NA lens.
This is achieved with an independent measure by exploiting the birefringence of the micropillar cavity~\cite{hilaire_accurate_2018}.
We thus obtain a free-space first lens brightness of \qty{46.8(2.5)}{\percent}.
Assuming that the first lens brightness of the micropillar is unaltered by the pigtailing process, we infer a micropillar-to-fiber coupling of \qty{75(5)}{\percent} from the measured fibered brightness of the pigtailed device, the transmission of the fiber splice between high-NA and standard fibers (\qty{90(2)}{\percent}), and the transmission of the filtering system (\qty{66(2)}{\percent}).
The measured coupling efficiency is comparable but somewhat higher than the expected value of \qty{71}{\percent} predicted by numerical simulations for our gap (\qty{3.5}{\micro\metre}) and micropillar size (\qty{2.8}{\micro\metre}).
This discrepancy can be attributed to imperfections on the measured micropillar size or to an overestimation of the losses of the splice and the filtering setup.

Next, we investigate the quality of the source by measuring the single-photon purity and indistinguishability of the emitted photons.
In Fig.~\ref{fig4}c we show the pulsed second-order intensity autocorrelation function, exhibiting strong photon antibunching at zero time delay with g\textsuperscript{(2)}(0)~=~\qty{1.3(1)}{\percent}.
In Fig.~\ref{fig4}d we also show the outcome of Hong-Ou-Mandel (HOM) interferometry which gives a two-photon HOM visibility of V\textsubscript{HOM}~=~\qty{95.0(1)}{\percent}.
After correction for the multi-photon events and the small imperfections of the interferometer~\cite{ollivier_hong-ou-mandel_2021}, we obtain an indistinguishability M~=~\qty{97.5(1)}{\percent}.
This value represents a significant improvement compared to the indistinguishability of the best pigtailed source reported so far of \qty{90(5)}{\percent}~\cite{snijders_fiber-coupled_2018}.
We note that the indistinguishability has increased from M~=~\qty{94.3(1)}{\percent} in the free-space characterization setup, which we attribute to the lower working temperature of \qty{2.4}{\kelvin} for the cryocooler hosting the fiber-coupled device compared to \qty{5}{\kelvin} for the low-vibration cryocooler used for the free-space characterization.
At lower temperatures, the residual contribution of the phonon sidebands is further reduced~\cite{reigue_probing_2017,grange_reducing_2017}.

Once the source is set to its best working conditions, we test its stability over ten hours of continuous operation.
In this measurement, only the excitation laser power is actively stabilized, with no stabilization of the fiber polarization nor of the bias voltage.
In Fig.~\ref{fig5}a,b we show the corresponding time traces of the single-photon rate and the photon indistinguishability.
Their distribution is depicted in Fig.~\ref{fig5}c,d, characterized by a relative standard deviation of \qty{2.82}{\percent} for the single-photon rate and of \qty{0.69}{\percent} for the indistinguishability.
We attribute the observed deviations from the maximum value to charge noise from using Be as p-type dopant.
The diffusivity of Be in GaAs reduces the quality of the p-i-n junction and its effect on the reduction of charge noise.
Such perturbations bring the system temporarily out of the optimal resonance condition, affecting both parameters at the same time.
This issue can be solved by using C doping, which is a more stable p-type option~\cite{malik_carbon_1988}.

\subsection{Discussion and conclusion}

\begin{figure*}[ht]
    \includegraphics[width=17.2cm]{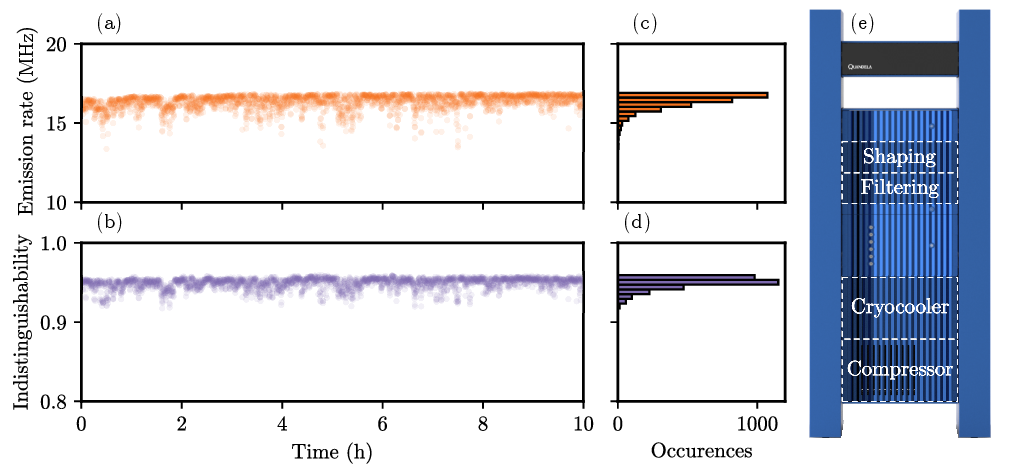}
    \caption{\label{fig5}Stability performance.
    (a,b) Stability over ten hours of the single-photon rate at the fiber output and photon indistinguishability and (c,d) corresponding histograms.
    (e) Standalone single-photon source system with interconnected modules dedicated to cryogenics, excitation and collection.}
\end{figure*}

In conclusion, we have demonstrated a reliable and efficient method for the pigtailing of micropillar-based single-photon sources that enables easy integration into compact cryocoolers.
We have described the fiber-coupling process and its main challenges, and presented the results of reflectivity measurements to check the coupling stability over time.
The pigtailed system shows record performances in fibered brightness (\qty{20.8}{\percent}) and indistinguishability (\qty{97.5}{\percent}).
These results show a large improvement over the previous state of the art, with comparable metrics with respect to the best single-photon sources in general.
This is achieved in a fully deterministic device, including the QD controllably coupled spatially and spectrally to a confined cavity mode, and the fiber accurately aligned on said cavity mode.
Moreover, the performances are shown to be stable over hours of uninterrupted operation and the coupling is maintained intact for months in different cryocoolers.

To further improve the efficiency of the system there are several parallel approaches that can be followed.
While LA excitation provides an inherent stability, the resulting occupation probability of the QD target state is limited to \qty{85}{\percent}~\cite{thomas_bright_2021} in the best conditions and it is estimated to be \qty{70}{\percent} for the larger detuning of \qty{-0.8}{\nano\metre} used here.
Other schemes such as resonant or swing-up~\cite{bracht_swing-up_2021} excitation allow to increase the occupation probability to near unity.
Regarding the pigtailing process, it is possible to engineer the resin stamp on the fiber tip to approach it even closer to the micropillar without contact.
Another area of improvement is the optimal mode-matching of the micropillar and fiber, which can be greatly improved by fabricating micropillars with better adjusted effective diameters.

The robust fiber pigtailing solution presented in this work demonstrates great potential for applications of single-photon sources beyond academic laboratories.
All the components needed for operating our single-photon sources are mounted in a rack system as depicted in Fig.~\ref{fig5}e.
This includes the compressor and the cryocooler hosting the pigtailed single-photon source, the pulsed laser with its spectral shaping module, the optics for filtering the single-photon emission, and a module for single-photon purity and indistinguishability measurements.
Moreover, it is possible to host multiple sources within the same cryocooler, which reduces the energy requirements compared to running a separate cryocooler for each source.
We estimate a total power consumption of \qty{3}{\kilo\watt} during operation.
The stability of the system can be further improved by implementing an active stabilization of the fiber polarization and an optimization of the bias voltage, especially for longer operation periods~\cite{maring_one_2023}.
As such, the single-photon source can be operated as a genuine turn-key system.

\subsection*{Acknowledgments}
This work has been partly funded by the European Commission as part of the EIC accelerator program under the grant agreement 190188855 for the SEPOQC project, by the Horizon-CL4 program under the grant agreement 101135288 for the EPIQUE project, by the European Union's Horizon 2020 Research and Innovation Programme QuDot-Tech under the Marie Skłodowska-Curie grant agreement 861097, by the Plan France 2030 through the projects ANR-22-PETQ-0011 and ANR-22-PETQ-0013, and by the French RENATECH network.

\newpage
\pagestyle{empty}
\bibliography{bibliography}{}
\bibliographystyle{ieeetr}

\end{document}